\documentclass[%
reprint,
superscriptaddress,
 amsmath,amssymb,
 aps,
prl,
]{revtex4-2}

\usepackage{graphicx}% Include figure files
\usepackage{dcolumn}% Align table columns on decimal point
\usepackage{bm}
\usepackage{float}% bold math
\begin{document}

%\preprint{APS/123-QED}

\title{Density-tuned isotherms and dynamic change at phase transition in a gate-controlled superconducting system}

\author{Shamashis Sengupta}
\email[]{sengupta@ijclab.in2p3.fr}
\affiliation{Universit\'{e} Paris-Saclay, CNRS/IN2P3, IJCLab, 91405 Orsay, France}

\author{Miguel Monteverde}
\affiliation{Universit\'e Paris-Saclay, CNRS, Laboratoire de Physique des Solides, 91405, Orsay, France}

\author{Anil Murani}
\affiliation{Universit\'e Paris-Saclay, CNRS, Laboratoire de Physique des Solides, 91405, Orsay, France}

\author{Claire Marrache-Kikuchi}
\affiliation{Universit\'{e} Paris-Saclay, CNRS/IN2P3, IJCLab, 91405 Orsay, France}

\author{Andr\'es F. Santander-Syro}
\affiliation{Universit\'e Paris-Saclay, CNRS,  Institut des Sciences Mol\'eculaires d'Orsay, 91405, Orsay, France}

\author{Franck Fortuna}
\affiliation{Universit\'e Paris-Saclay, CNRS,  Institut des Sciences Mol\'eculaires d'Orsay, 91405, Orsay, France}


\begin{abstract}
Two-dimensional electron gases in SrTiO$_3$-based heterostructures provide a platform to study the real-time evolution of the macroscopic state with a variation of the carrier density, and the impact of structural properties on the emergence of the superconducting state. We have explored the isothermal evolution of the electron gas in AlO$_x$/SrTiO$_3$ by measuring the variation of resistance with continuous gate-voltage-controlled tuning of its carrier density. It is seen that condensation of the ordered phase leads to non-monotonic isotherms within the superconducting dome. The timescale for dynamic change following changes in gate voltage is measured across the phase transition. It is found to be tens of seconds near the onset of superconductivity, significantly larger compared to the normal state. Such a large timescale governing the kinetics of the phase transition presumably arises from the strong impact of structural defects and distortions of the substrate on the development of superconducting islands.
\end{abstract}

\maketitle

%\tableofcontents

%\section{\label{sec:level1}First-level heading:\protect\\ The line
%break was forced \lowercase{via} \textbackslash\textbackslash}

\section{I. Introduction}

The phenomenon of superconductivity in SrTiO$_3$ has many interesting aspects. As a function of doping, it exhibits a superconducting dome \cite{schooley,behnia} in the phase diagram. The carrier densities in superconducting SrTiO$_ 3$ are extremely low, due to which the nature of the microscopic pairing mechanism remains unclear. The interaction of electrons with polar optical phonons have been explored in this context \cite{gorkov,enderlein}. From the point of view of structural properties, SrTiO$_3$ undergoes a ferroelastic transition \cite{fleury,salje} from cubic to tetragonal symmetry at 105 K. At very low temperatures, zero-point motion suppresses a ferroelectric transition making it a quantum paraelectric \cite{muller}. Understanding the role of the structural properties of SrTiO$_3$ on the emergence of the superconducting state is a topic of general interest.

A pertinent development in recent years is the realization of superconducting two-dimensional electron gases (2DEGs) in SrTiO$_3$-based heterostructures \cite{reyren}. The carrier density of such 2DEGs is tunable with an electrostatic gate voltage, leading to the observation of a superconducting dome \cite{caviglia, biscaras, ilani, hwang, dikin}. In low-dimensional superconductors, the presence of disorder is predicted \cite{dubi} to lead to spatial fluctuations of the order parameter. Transport experiments on SrTiO$_3$-based heterostructures suggest that the superconducting state can be modelled as a set of percolating superconducting islands in a metallic background \cite{biscaras,pryds_superconductivity}. In these systems, the superconducting transition is usually studied by stabilizing the carrier density $n$ at certain fixed values of the gate voltage $V_g$, while continuously varying the temperature. In other words, the phase transition is induced along an isochore (Fig. 1a). In this work, our primary aim is to probe the development of the superconducting state as the carrier density is tuned with gate voltage under isothermal conditions. The objective is to observe the onset of the superconducting state with introduction of carriers, which in turn leads to the growth of superconducting islands. This process is likely to be influenced by the interaction of the electronic system with the substrate, which can be studied by measurements of dynamic change in the vicinity of the phase transition.

Our experiments were conducted on a superconducting 2DEG in the AlO$_x$/SrTiO$_3$ heterostructure. In Section II, we will describe the realization of the heterostructure and characterization of its transport properties. Section III is concerned with measurements on the isothermal variation of resistance ($R$) with a continuous tuning of gate voltage ($V_g$) close to the maximum critical temperature ($T_m$) of the superconducting dome (Fig. 1a). The onset of superconductivity is marked by the development of non-monotonic $R-V_g$ isotherms. The application of a gate voltage results in a finite electric field \cite{field} across the bulk SrTiO$_3$ dielectric. This leads to the observation of non-equilibrium effects associated with the onset of superconductivity. In our experiments we have measured the stabilization time ($\tau$) for attainment of steady state following changes in the gate voltage. The characteristic timescale of evolution is found to be much larger in the superconducting state compared to the normal state. These results will be presented in Section IV.

\begin{figure*}[ht]
%\begin{center}
\includegraphics[width=178mm]{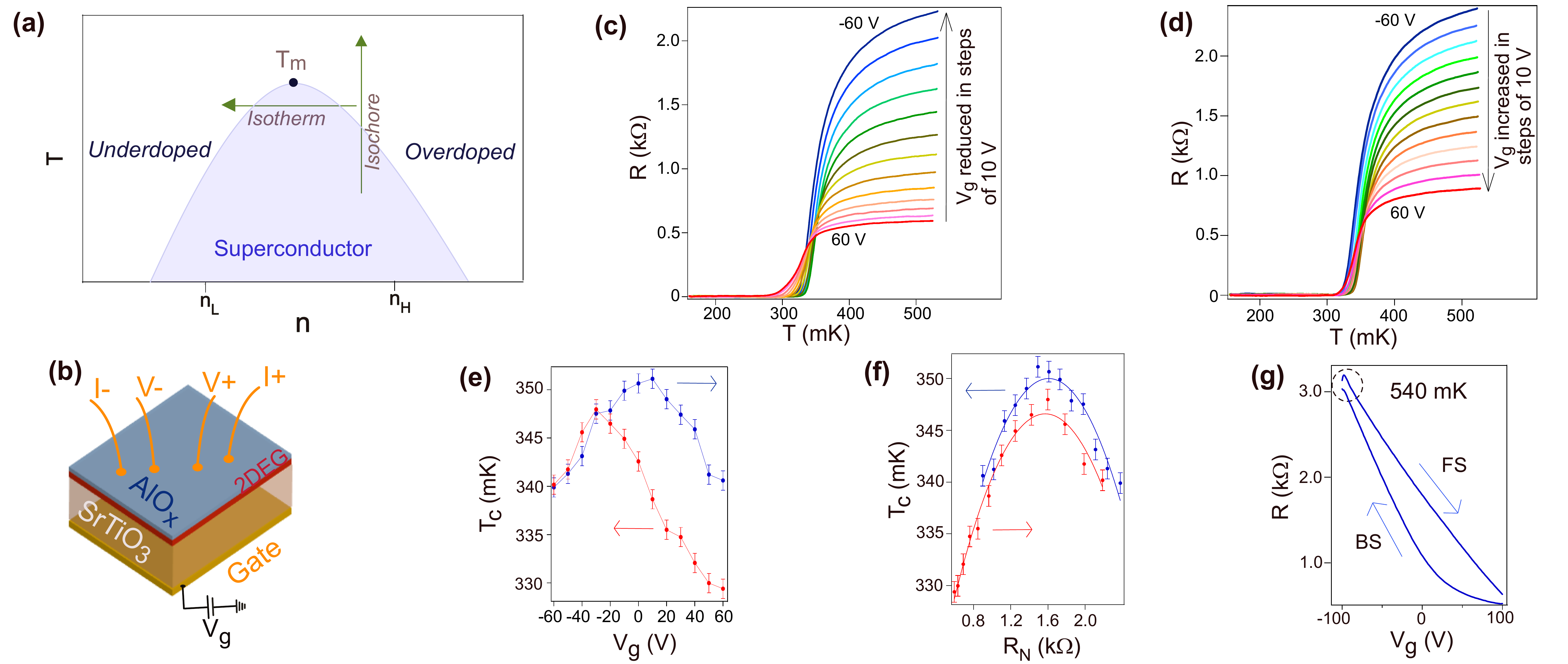}
\caption{\textbf{(a)} Schematic of a superconducting dome phase diagram plotted as a function of temperature ($T$) and carrier density ($n$). The range of carrier densities accessible in our experiments (from $n_L$ to $n_H$) lies within the superconducting dome. \textbf{(b)} Schematic diagram of the heterostructure with wire-bonded contacts. \textbf{(c, d)} Four-probe resistance ($R$) is measured as a function of temperature ($T$) for different values of gate voltage $V_g$. The dc current applied is 20 nA. \textbf{(e)} The critical temperature $T_c$ plotted as a function of $V_g$. Arrows indicate the chronological order in which the data, presented in (c) and (d), pertaining to these points was acquired. \textbf{(f)} $T_c$ plotted as a function of $R_N$ (the resistance at 500 mK). Solid lines are guides to the eye. \textbf{(g)} Resistance ($R$) measured with a continuous variation of $V_g$ (at a rate of 0.28 V/s). The current applied is 20 nA.}
%\end{center}
\end{figure*}

\section{II. Characterization of transport properties}

The AlO$_x$/SrTiO$_3$ heterostructure was prepared by using a method described in Refs.~\citenum{sengupta, rodel}. A (001)-oriented SrTiO$_3$ crystal (measuring 5 mm x 5 mm x 0.5 mm) was placed inside an UHV (ultrahigh vacuum) chamber. An annealing step was performed to clean the surface contamination. The sample was heated to $600^{\circ}$C for 1 minute. Then the temperature was allowed to reduce, and 2 nm of Al was evaporated using a Knudsen cell at a temperature higher than room temperature ($200^{\circ}$~C). The rate of deposition of Al was 0.002 nm/s. This produces an AlO$_x$/SrTiO$_3$ heterostructure with a conducting two-dimensional electron gas (2DEG) at the interface. Al removes O atoms from the surface of SrTiO$_3$ and is transformed into the insulating AlO$_x$ capping layer. Oxygen vacancies on the surface of SrTiO$_3$ lead to the creation of the 2DEG by the doping of Ti 3d levels  \cite{rodel, andres}.  The  backplane of the SrTiO$_3$ crystal was glued to a copper plate with conductive silver paint to serve as the gate electrode. The resistance of the SrTiO$_3$ substrate between the surface with 2DEG and the gate electrode was greater than 100 G$\Omega$.

Transport measurements were carried out by contacting the 2DEG with ultrasonic wire-bonding (Fig. 1b). First, the four-probe resistance ($R$) was measured as a function of the temperature ($T$) for different values of the gate voltage ($V_g$) allowing us to determine the $V_g$ corresponding to the maximum critical temperature $T_m$ of the dome. Fig. 1c shows the results when $V_g$ is changed from 60 V to -60 V (in steps of 10 V), corresponding to a progressive reduction of carrier density.  Fig. 1d shows the set of curves for the reverse direction of change in $V_g$. (See Supplemental Material \cite{supp} for more results.) The critical temperature ($T_c$) is defined to be the temperature where the slope of the superconducting resistance drop ($\frac{dR}{dT}$) is maximum. The variation of $T_c$ with $V_g$ is shown in Fig. 1e. The domes for decreasing and increasing directions of gate voltages show a hysteresis with the maximum $T_c$ appearing at different values of $V_g$. Hysteresis with changes in gate voltage may result from the presence of charged defects (mobile oxygen vacancies or charge traps) in the SrTiO$_3$ dielectric \cite{hwang} or the time-dependent response of ferroic domains of the substrate to an applied electric field \cite{pesquera}. Such factors lead to different doping configurations for forward and backward sweeps of $V_g$. In Fig. 1f, the superconducting dome is represented as the variation of $T_c$ with respect to the high-temperature resistance at 500 mK ($R_N$). The maximum of $T_c$ occurs for $R_N$ $\thicksim$ 1.5 - 1.6 k$\Omega$ for both directions of change of $V_g$. This confirms that the relevant parameter for identifying the maximum critical temperature of the superconducting dome is indeed the carrier density. The result of a continuous tuning of $V_g$ (at a sweep rate of 0.28 V/s), first in backward sweep (BS) and then in forward sweep (FS), is shown in Fig 1g. In their study of the electron trapping mechanism in LaAlO$_3$/SrTiO$_3$, Yin et al. \cite{hilgenkamp} observed that electromigration and clustering of oxygen vacancies may give rise to an evolution of the 2DEG resistance for several seconds after the $V_g$ has been stabilized following a ramp. We observe a similar effect in Fig. 1g (marked by a dashed circle). $R$ increases with decreasing $V_g$ during the BS, and continues to increase for a few seconds even after the $V_g$ sweep direction is changed to FS. For all measurements reported in this article, the sheet resistance (per square) can be obtained by multiplying the four-probe resistance by a factor of 3.2 (calculated from the geometry of the contacts). The Hall carrier densities determined from magnetic field sweeps up to 1 T at 100 mK are 1.9$\times$10$^{13}$ cm$^{-2}$, 2.4$\times$10$^{13}$ cm$^{-2}$ and 3.0$\times$10$^{13}$ cm$^{-2}$  for gate voltages of -80 V, 0 V and 80 V respectively.

\section{III. Tuning of carrier density under isothermal condition}

The result of isothermal tuning of carrier density is presented in Fig. 2.  The gate voltage is tuned first in BS from 100 V to - 100 V (Fig. 2a), then in FS from -100 V to +100 V (Fig. 2b) at a sweep rate of 0.28 V/s. At 395 mK, well above the onset of the macroscopic superconducting phase, we see the expected metallic behaviour of monotonic decrease (increase) in $R$ with increase (decrease) in carrier density. At 325 mK, the evolution of $R$ becomes non-monotonic with three distinct regions emerging (marked I, II, III). In region II, $\frac{dR}{dn}$ is positive. Here, a reduction of the  carrier density is accompanied by a reduction of the resistance. This non-monotonicity of the isotherm can be interpreted as a signature of the development of the superconducting phase.

A prototypical system with a superconducting dome (Fig. 2c) is composed of weakly interacting carriers in the overdoped regime (with large carrier density $n$). As $n$ is progressively reduced at constant temperature (below $T_m$ of the dome), first there is an augmentation of the mean-field superconducting gap followed by a gradual loss of phase coherence towards lower densities in the underdoped regime. In Fig. 2c, the line $T_{MF}$ denotes schematically the mean-field transition temperature and $T_\theta$ denotes the upper bound on the phase ordering temperature (following the theoretical work by Emery and Kivelson \cite{kivelson}). In the overdoped (OD) and underdoped (UD) regimes, outside the superconducting dome, the system is expected to show the characteristic of a normal conductor with negative $\frac{dR}{dn}$. This corresponds to regions I and III of the $R-n$ isotherm for $T<T_m$. At intermediate densities, due to the formation of finite superfluid density within the superconducting dome, there is a deviation from this behaviour marked by a sharp drop in resistance. This causes the change in slope with positive $\frac{dR}{dn}$ in region II of the isotherm.

\begin{figure}
%\begin{center}
\includegraphics[width=86mm]{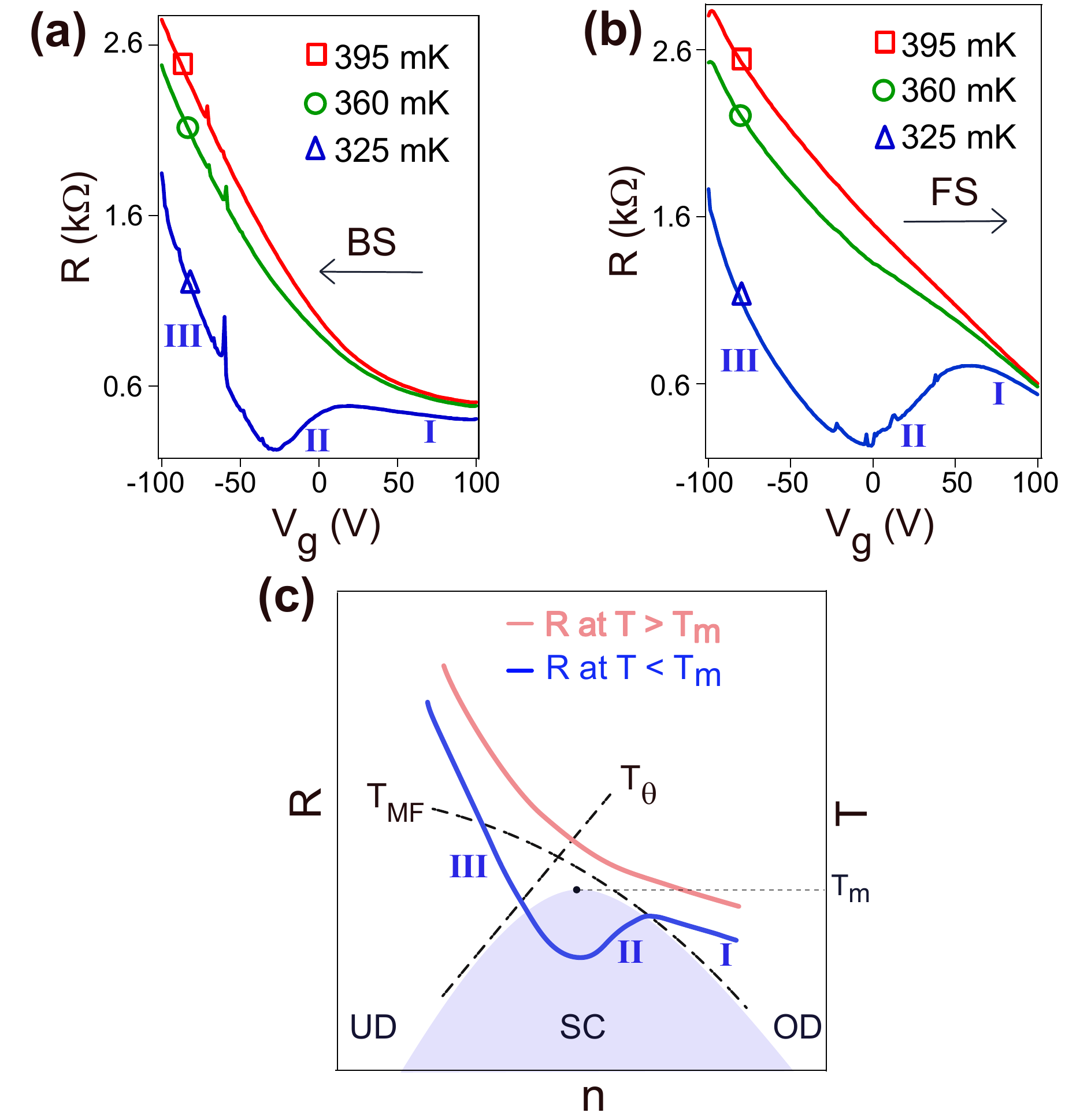}
\caption{\textbf{(a, b)} Variation of $R$ as a function of $V_g$ at three different temperatures is shown for BS \textbf{(a)} and FS \textbf{(b)} respectively. DC current applied was 20 nA. \textbf{(c)} Illustration of expected variation of $R$ (left-hand axis) as a function of carrier density $n$ along isotherms of a system exhibiting a superconducting dome. The phase diagram is represented with temperature $T$ along the right-hand axis.}
%\end{center}
\end{figure}

The carrier density $n$ varies monotonically with $V_g$. However, strictly speaking, an exact relation between the two can not be established because of hysteretic effects, as discussed earlier in Section II. The hysteresis is apparent in Figs. 2a and 2b, when we compare curves for BS and FS corresponding to the same temperature. The factors leading to the hysteresis, presumably the gate-voltage-induced movement of oxygen vacancies or ferroic domains, might be anticipated to influence the development of the superconducting state. These effects would be manifested in the kinetics of the nucleation and growth of superconducting islands. Experiments to study this phenomenon are described in the following section.

\section{IV. Dynamic change at the superconducting transition}

We will now discuss about the measurement of dynamic change in $R$  following a gate-induced change in carrier density. The gate voltage is ramped from high to low $V_g$ at a rate of 0.28 V/s, progressively reducing $n$, before it is stopped at a certain value of $V_g$ (where it is thereafter held steady). The resistance $R$ is then monitored with time $t$. This set of measurements is performed at 480 mK (Figs. 3a and 3b) and 325 mK (Figs. 3c and 3d), stopping the $V_g$ ramp at values of -70 V (BS1), -85 V (BS2) and -95 V (BS3). At 480 mK, $R$ becomes steady after a few seconds (Fig. 3b). This can be explained as an effect of the motion of charged defects within the bulk substrate (as mentioned earlier). Upon lowering the temperature to 325 mK, the dynamic change with time becomes quite different. Instead of an increase in $R$, we observe a larger reduction in $R$ (Fig. 3d). More importantly, the evolution takes place over a much longer time before an almost steady state is reached (Fig. 3d). At $V_g$=-70 V, even after 500 seconds of wait, $R$ still continues to reduce with time, showing that the time to reach a steady state is even longer. To quantify the characteristic timescales, we fit the data in Figs. 3b and 3d to the function $R(t)=R_0 +$  $\alpha e^{-t/\tau}$ $+$ $\beta t$ with $R_0$, $\alpha$, $\beta$ and $\tau$ as fit parameters. The term $\beta t$ accounts for the gradual change in resistance at large $t$.  The term $\alpha e^{-t/\tau}$ captures the sharp change in resistance just after $t$=0 with a characteristic time $\tau$. $\tau$ is found to be 5 s, 3 s, 3 s for $V_g$ values of -70 V, -85 V, -95 V respectively at 480 mK (Fig. 3b). At 325 mK (below $T_m$ of the superconducting dome), these increase to 68 s, 28 s and 19 s respectively (Fig. 3d). The increase in $\tau$ appears to be directly related to the development of the superconducting phase. To cross-check this inference, we repeated these measurements in presence of a magnetic field.

\begin{figure}
%\begin{center}
\includegraphics[width=86mm]{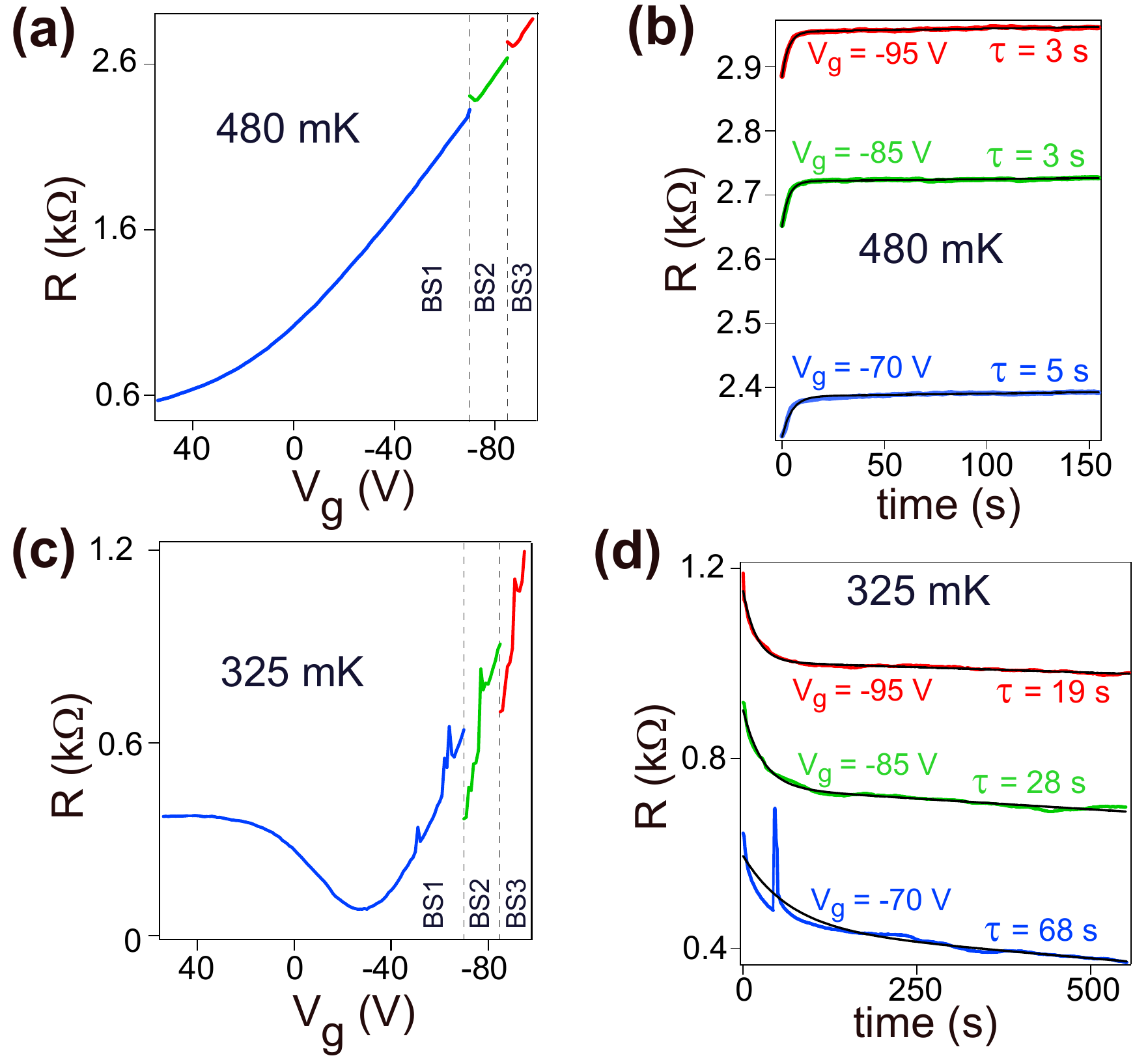}
\caption{\textbf{(a,b)} $V_g$ is swept at 0.28 V/s  at 480 mK temperature. \textbf{(a)} shows the ensuing change in $R$. Three sets of measurements are shown (BS1, BS2 and BS3). Dynamic change in $R$ upon stopping $V_g$ sweep is shown in \textbf{(b)}.  DC current applied is 20 nA. \textbf{(c,d)} The same experimental protocol described in (a,b) is followed, at 325 mK. Variation of $R$ during $V_g$ sweep is shown in \textbf{(c)}. Dynamic change in $R$ after stabilizing $V_g$ is shown in \textbf{(d)}.}
%\end{center}
\end{figure}

\begin{figure*}[ht]
\begin{center}
\includegraphics[width=178mm]{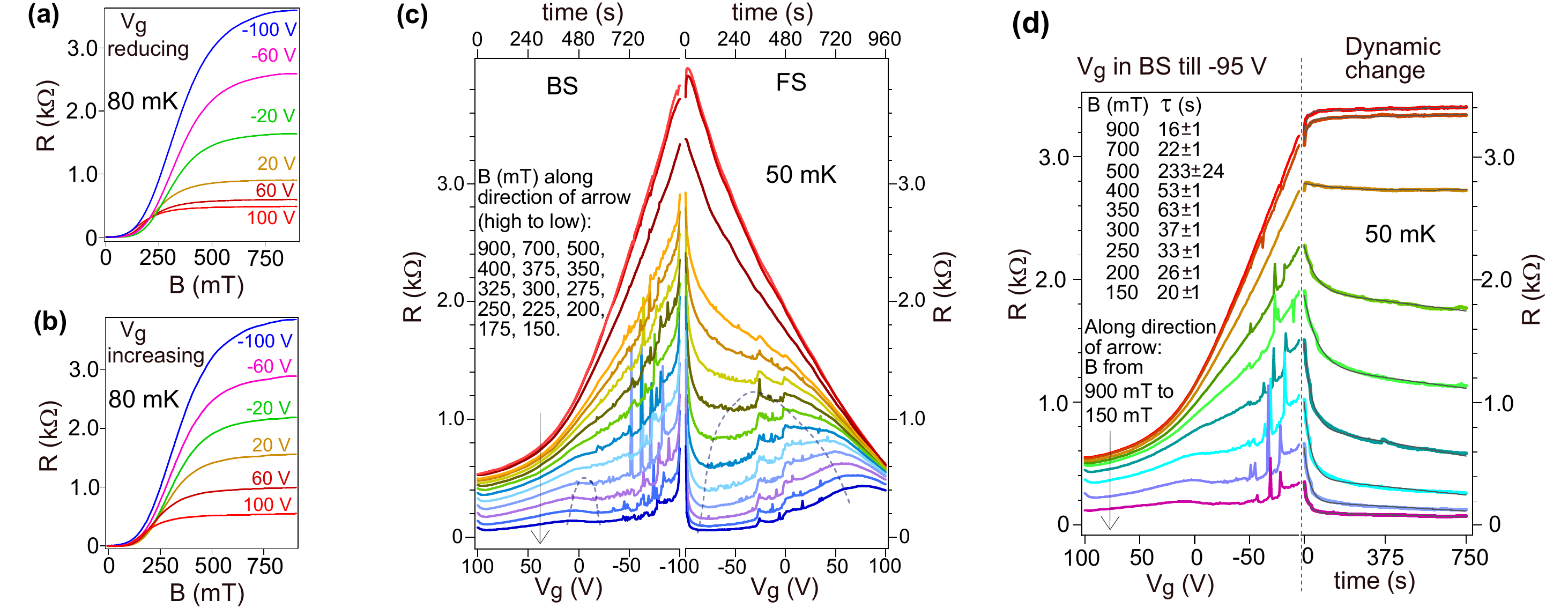}
\caption{\textbf{(a,b)} $R$ is measured as a function of magnetic field ($B$) for different values of $V_g$ at 80 mK. DC current applied is 20 nA in (a) and 100 nA in (b). \textbf{(c)} $R$ is measured as a function of $V_g$ for different values of $B$ at 50 mK. $V_g$ is first swept in back sweep (BS) and then in forward sweep (FS) at 0.21 V/s. Time increases from left to right. DC current applied is 50 nA. Dashed lines correspond to regions where $\frac{dR}{dn}$ is positive. \textbf{(d)} $V_g$ is swept at 0.21 V/s from 100 V to -95 V with progressive depletion of the electron gas at different values of $B$ (left of dotted line). Then $V_g$ is stabilized and dynamic change in $R$ is recorded with time (right of the dotted line). Solid lines are fits to the data. DC current used is 10 nA.}
\end{center}
\end{figure*}

The evolution of 2DEG resistance as a function of the magnetic field $B$ was measured for different fixed values of $V_g$ (Figs. 4a and 4b). We then measured the evolution of resistance with a continuous tuning of $V_g$ at different values of $B$, with the temperature at 50 mK. These isolines of constant magnetic field are shown in Fig. 4c. $V_g$ was first changed in BS from 100 V to -100 V, then in FS from -100 V to 100 V (at a rate of 0.21 V/s). A clear difference is observed between the curves for BS and FS. A rapid drop in $R$ is seen at the beginning of the FS, for field values below 400 mT. At low fields, a prominent region of positive $\frac{dR}{dn}$ appears (indicated by a dashed curve) during forward $V_g$ sweeps. This region is however barely visible in backward $V_g$ sweep. This difference, we may infer, is related to the kinetics of the superconducting state. The microscopic configuration of the superconducting electronic system evolves differently during the introduction of carriers (FS) as compared to the depletion of carriers (BS). The electronic system is likely to be composed \cite{biscaras, hwang_islands, pryds_superconductivity} of several superconducting islands. The degree of phase coherence between these islands determines the extent of macroscopic order. The rapid drop in $R$, seen at the beginning of the FS (Fig. 4c) for field values less than 400 mT, suggests the formation of large phase coherent superconducting droplets upon the reversal of sweep direction of $V_g$. (See Supplemental Material \cite{supp} for the outcome of varying $V_g$ sweep rates.) As we will discuss now, the development of superconducting islands may be strongly influenced by intrinsic properties of the SrTiO$_3$ dielectric.

The dynamic change of $R$ for attainment of steady state was measured by ramping $V_g$ down till -95 V (at 0.21 V/s) and recording the ensuing temporal evolution (Fig. 4d). At high fields, $R$ increases for a certain time ($\tau$ of 16 s at 900 mT and 22 s at 700 mT) before stabilizing at a steady value. At 500 mT, a different behaviour sets in. There is a momentary increase in $R$ followed by a slow drop over an extended period of time. $\tau$ is estimated to be 233 s. (See Supplemental Material \cite{supp} for further discussions.) At lower fields, $R$ drops by larger amounts when the $V_g$ sweep is stopped at -95 V. The noticeable difference of dynamic change behaviour at high and low fields can be explained by the fact that the low field behaviour is related to the kinetics of superconducting islands. The resistance drop corresponds to the condensation or the increase in size of such islands. The estimated times of several seconds are difficult to explain in terms of a purely electronic mechanism. Transient superconductivity seen in different systems with supercritical current pulses are associated with relaxation times of a few nanoseconds or few hundred picoseconds \cite{pals,frank,tinkham}. The large values of $\tau$ in our experiment are explained much better by taking into account the properties of the SrTiO$_3$ substrate hosting the 2DEG and their influence on the superconducting transition. There are two aspects of the structural properties of SrTiO$_3$ which are of relevance here - oxygen vacancies and ferroic domains. In the following paragraphs, we will discuss about the possible impact of these features on the development of superconducting islands.

It is known that relaxation processes due to movement of oxygen vacancies inside SrTiO$_3$, upon the application of an electric field \cite{hilgenkamp} or ultra-violet radiation \cite{meevasanaa}, may last for several seconds or even minutes. The motion of such charged defects under an electric field in our system can account for the observed values of $\tau$ in the high temperature (Fig. 3b) or high-field (Fig. 4d) normal state. The moving defects may provide a fluctuating potential landscape for the electrons, affecting the nucleation of superconducting droplets or the expansion of phase coherent superconducting islands. The inference is that nucleation and growth of superconducting islands are facilitated by dynamic changes in the structural properties at the surface of SrTiO$_3$. A small increase in the fraction of superconducting electrons will result in a significant reduction of the resistance. Accordingly, the dynamic change in resistance will be larger in magnitude and appreciable over a longer span of time with the onset of superconductivity.

Apart from oxygen vacancies, ferroic domains within SrTiO$_3$ are also likely to influence the development of the superconducting state. The impact of such domains on the phenomenon of electronic conduction has been investigated in previous studies. Kalisky et al. \cite{kaliskystructure} studied the LaAlO$_3$/SrTiO$_3$ interface electronic system with scanning superconducting quantum interference device (SQUID) microscopy and observed that the local conductivity is modified by the tetragonal domain structure of SrTiO$_3$. The current flow at low temperatures occurred preferably in narrow paths oriented along the crystallographic axes. These current-carrying channels were embedded in a less conductive background. Noad et al.  \cite{moler} used scanning SQUID susceptometer to probe the superconductivity in thin Nb-doped SrTiO$_3$ layers embedded in undoped SrTiO$_3$. They observed spatial variations of the critical temperature in a manner similar to the configuration of structural twin domains. Honig et al. \cite{ilanistructure} imaged the LaAlO$_3$/SrTiO$_3$ system with a scanning charge detector. They observed tetragonal domains in SrTiO$_3$ and the movement of domains with the variation of gate voltage. Pai et al. \cite{levy} conducted transport experiments on narrow channels in LaAlO$_3$/SrTiO$_3$ and concluded that ferroelastic domain boundaries may have an important role in the formation of the superconducting state.

Pesquera et al. \cite{pesquera} demonstrated that polar domains in SrTiO$_3$ under an applied electric field exhibit relaxation times of several tens of minutes (at 36 K temperature). This suggests that the large $\tau$ observed in our experiments might be related to the relaxation of structural domains within the dielectric and their influence upon the superconducting state. This can be understood in the following way.  Under the application of an electric field across the SrTiO$_3$ dielectric, the ferroic domains undergo a relaxation process over exceedingly large timescales ($\tau_f$) to attain a steady state configuration. Such processes affect the properties of the conducting 2DEG at the surface. Accordingly, the resistance (or conductance) of the system undergoes an evolution over long timescales (which is the quantity $\tau$ determined from our measurements). Electrical transport measurements are not capable of determining $\tau_f$. $\tau$ need not be the same as $\tau_f$, since the impact of domain relaxation on the conducting properties of the 2DEG would depend on the specific electronic state of the 2DEG as well. $\tau$ can vary with temperature and magnetic field depending upon the microscopic structure of the 2DEG. At temperature or magnetic fields low enough for superconductivity to have appeared, the system consists of both normal metallic regions and superconducting islands. Certain metallic regions appear to be susceptible to undergo a transition into the superconducting state upon small changes in the local environment. Such a transition might be energetically favourable in some parts of the electronic system, and is guided by the interaction of the 2DEG with the ferroic domains. The increase in size and number of superconducting islands leads to new features that are not visible at higher temperatures or magnetic fields (when no superconductivity exists).  While such changes do not lead to substantial dynamic change in resistance in the purely normal state (when $\tau$ is small), these do affect the resistance prominently in the superconducting state (when $\tau$ is large).

Spike-like features are visible in the data shown in Figs. 3d, 4c and 4d. These can result from sudden transformations in the arrangement of trapped charges or ferroic domains within the substrate, which influence the properties of the interface electronic system. It is noteworthy that such spike-like features in resistance are visible only at temperature and magnetic field values low enough for the onset of superconductivity to have taken place. These are not visible in the normal state, corroborating our inference that the 2DEG in AlO$_x$/SrTiO$_3$ is much more sensitive to structural defects or deformations of the substrate in the superconducting state than in the normal state.

\section{V. Conclusions}

In conclusion, we have explored the superconducting dome in an AlO$_x$/SrTiO$_3$ heterostructure by isothermally varying its carrier density with a gate voltage $V_g$ and measuring the evolution of resistance $R$. The $R$-$V_g$ isotherms are observed to be non-monotonic below the maximum critical temperature of the superconducting dome due to the non-monotonic variation of superfluid density as a function of carrier density. Dynamic change of tens of seconds is observed following a change of gate voltage close to the phase transition, highlighting the sensitivity of the superconducting state to structural properties of the underlying substrate, which might be the distribution of oxygen vacancies or ferroic domains. Due to such large timescales, the kinetics of the real time growth of superconducting islands becomes visible in dc transport measurements. Previous experiments on SrTiO$_3$-based heterostructures highlighted the fact that the properties of the superconducting state are deeply influenced by inhomogeneities. Based on critical current measurement, Prawiroatmodjo et al. \cite{pryds_superconductivity} concluded that the superconducting system in LaAlO$_3$/SrTiO$_3$ can be modelled as a Josephson junction array of weakly coupled superconducting domains, with the inhomogeneities arising from the distribution of oxygen vacancies or tetragonal domain boundaries of SrTiO$_3$. From transport studies on nanoscale conducting channels, Pai et al. \cite{levy} suggested that ferroelastic domain boundaries play an important role in the phenomenon of superconductivity. We have shown that dynamic change measurements at the onset of superconductivity support these observations and provide a useful method of investigating the interaction between the electronic system and structural properties of the substrate.

In a recent experiment, Kremen et al. \cite{kalisky} used scanning probe microscopy to look into order parameter fluctuations in disordered superconductors. Such experimental techniques can provide new insights into the kinetics of the superconducting transition in SrTiO$_3$-based heterostructures and can be interesting topics for further research.  The experimental protocols used in our work may be applied more generally to study the dynamic change behaviour in different other superconducting systems, notably ultrathin materials \cite{mos2_science2, tise2, graphene} with gate-tunable superconductivity.

\section{Acknowledgments}
The authors thank R. Deblock, H. Bouchiat, S. Gu\'eron and P. Senzier for help during the experiments. We thank D. Petrov for insightful discussions. This work was supported by public grants from the French National Research Agency (ANR), project CP-Insulators No. ANR-2019-CE30-0014-03.

%\nocite{*}


\begin{thebibliography}{99}
\bibitem{field}
The electric field corresponding to a gate voltage $V_g$ is $V_g$/$t$, where $t$ is the thickness (0.5 mm) of the SrTiO$_3$ crystal.
\end{thebibliography}

%%%%%%%%%% Merge with supplemental materials %%%%%%%%%%
\pagebreak
\widetext
\begin{center}
\textbf{\large Supplemental Material}
\end{center}
%%%%%%%%%% Merge with supplemental materials %%%%%%%%%%
%%%%%%%%%% Prefix a "S" to all equations, figures, tables and reset the counter %%%%%%%%%%
\setcounter{equation}{0}
\setcounter{figure}{0}
\setcounter{table}{0}
\setcounter{page}{1}
\makeatletter
\renewcommand{\theequation}{S\arabic{equation}}
\renewcommand{\thefigure}{S\arabic{figure}}
\renewcommand{\bibnumfmt}[1]{[S#1]}
\renewcommand{\citenumfont}[1]{S#1}
%%%%%%%%%% Prefix a "S" to all equations, figures, tables and reset the counter %%%%%%%%%%

%\section{1. S\lowercase{ample fabrication}}

%\section{2. S\lowercase{uperconducting transition at a gate voltage of -100 }V}

%\begin{figure}[ht!]
%\begin{center}
%\includegraphics[width=60mm]{Sup_Comp_m100V.pdf}
%\caption{Variation of resistance ($R$) as a function of temperature ($T$) at a gate voltage of -100 V. The current applied was 20 nA. The critical temperature (defined as the point of maximum slope $\frac{dR}{dT}$) is 324 mK.}
%\end{center}
%\end{figure}

\section{1. S\lowercase{uperconducting transition at a gate voltage of} -100 V}

We have presented data in the main text (Fig. 3d) of a significant reduction in resistance with time when the gate voltage sweep is stopped at $V_g$ values of -70 V, -85 V and -95 V. This measurement was done at 325 mK. The resistance measured is finite because this temperature is still above the temperature at which resistance becomes zero. The variation of resistance showing the full superconducting transition as a function of temperature, at a fixed gate voltage of -100 V, is shown in Fig. S1.

\begin{figure}[h]
\begin{center}
\includegraphics[width=80mm]{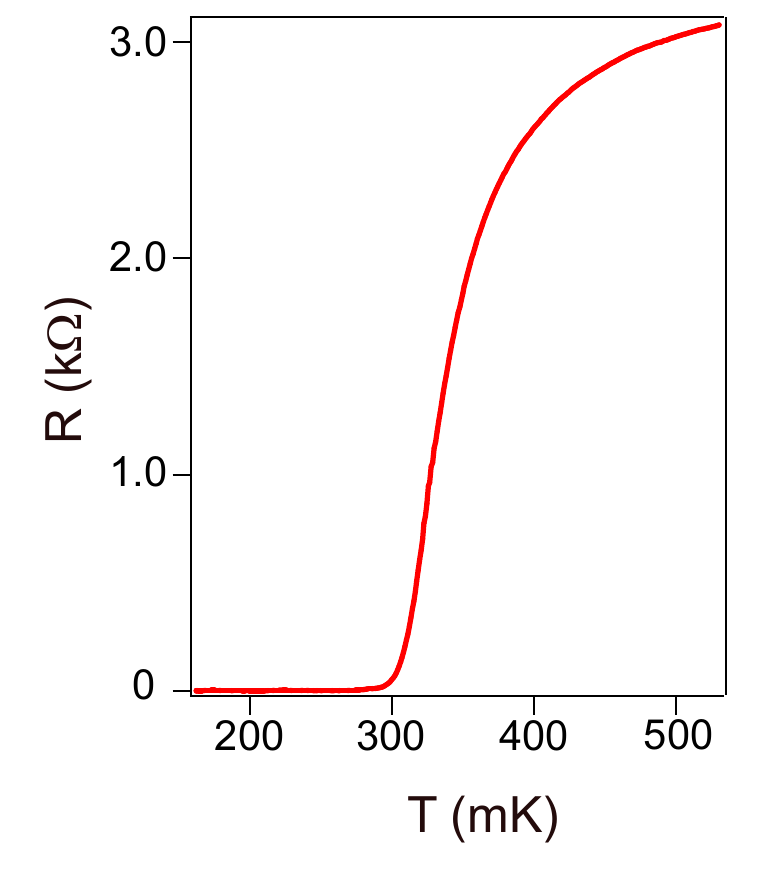}
\caption{Variation of resistance ($R$) as a function of temperature ($T$) at a gate voltage of -100 V. The current applied was 20 nA. The critical temperature (defined as the point of maximum slope $\frac{dR}{dT}$) is 324 mK.}
\end{center}
\end{figure}

\begin{figure}[h]
\begin{center}
\includegraphics[width=80mm]{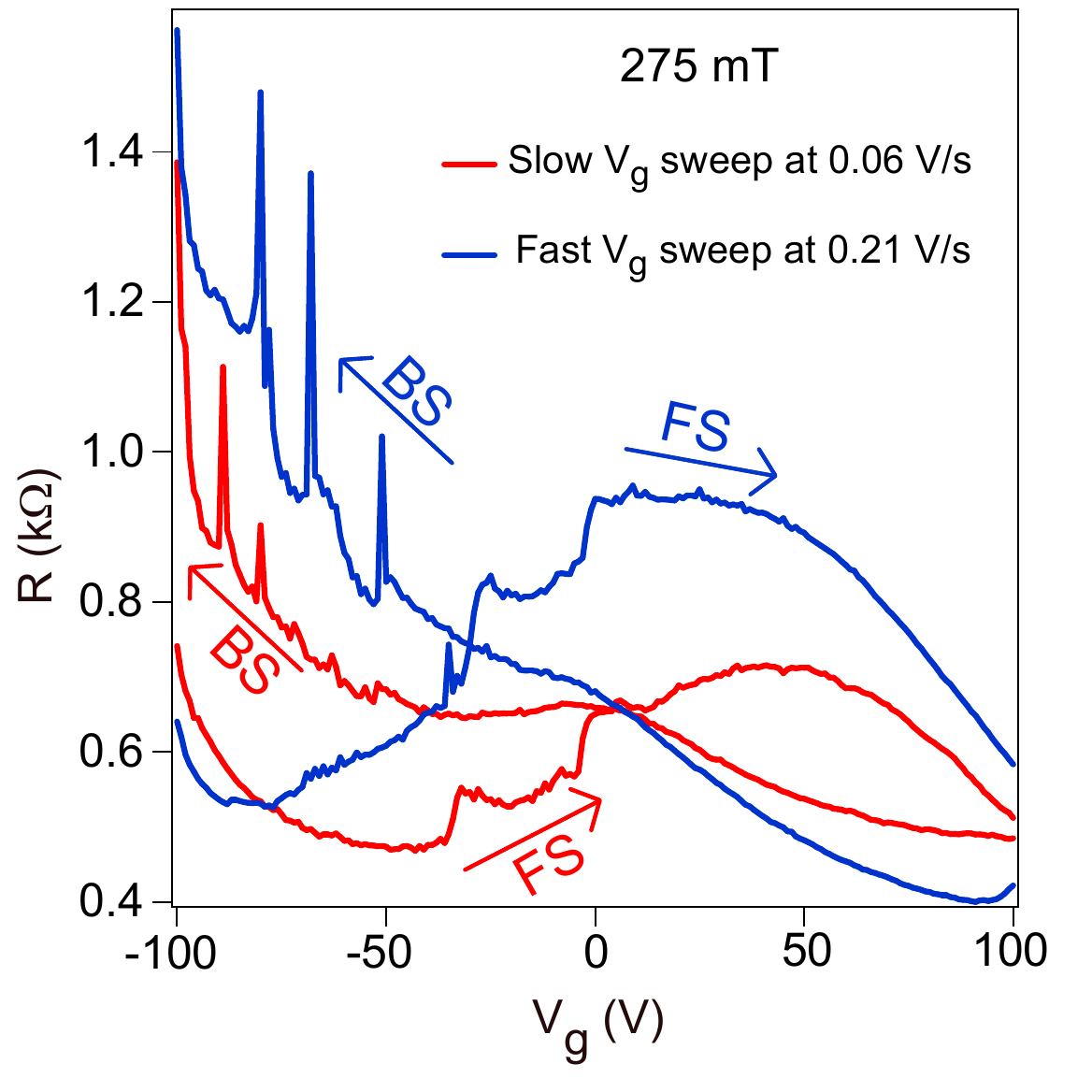}
\caption{Variation of resistance ($R$) with gate voltage ($V_g$) sweep. The temperature is 50 mK and applied magnetic field is 275 mT. The experimental protocol is the following. $V_g$ is ramped from 0 V to 100 V (not shown). Then, it is reduced in backsweep (BS) from 100 V to - 100 V. $V_g$ is held at -100 V for two minutes and then swept back to 100 V in forward sweep (FS). This set of measurements is carried out for two different $V_g$ sweep rates. The DC current used is 10 nA.}
\end{center}
\end{figure}

\section{2. T\lowercase{he result of variation of gate voltage sweep rate on the evolution of resistance}}

In the main text, we identified two possible causes for the large timescales in dynamic change behaviour of the resistance $R$ following a continuous change of the gate voltage $V_g$ near the onset of superconductivity. These were the motion of charged defects (primarily oxygen vacancies) within the bulk dielectric, and the relaxation of ferroic domains in the substrate. Both of these are known to exhibit dynamics lasting several seconds. Given this explanation, one can expect that a variation of the sweep rate will result in visible changes in the evolution of resistance. This experiment is shown in Fig. S2. $V_g$ was swept in both back sweep (BS) and forward sweep (FS) at 50 mK temperature with an applied magnetic field of 275 mT. Two different sweep rates were used: 0.21 V/s (fast) and 0.06 V/s (slow). The overall variation of $R$ is clearly less for the slower $V_g$ sweep in comparison to the faster one.

\begin{figure}[h]
\begin{center}
\includegraphics[width=80mm]{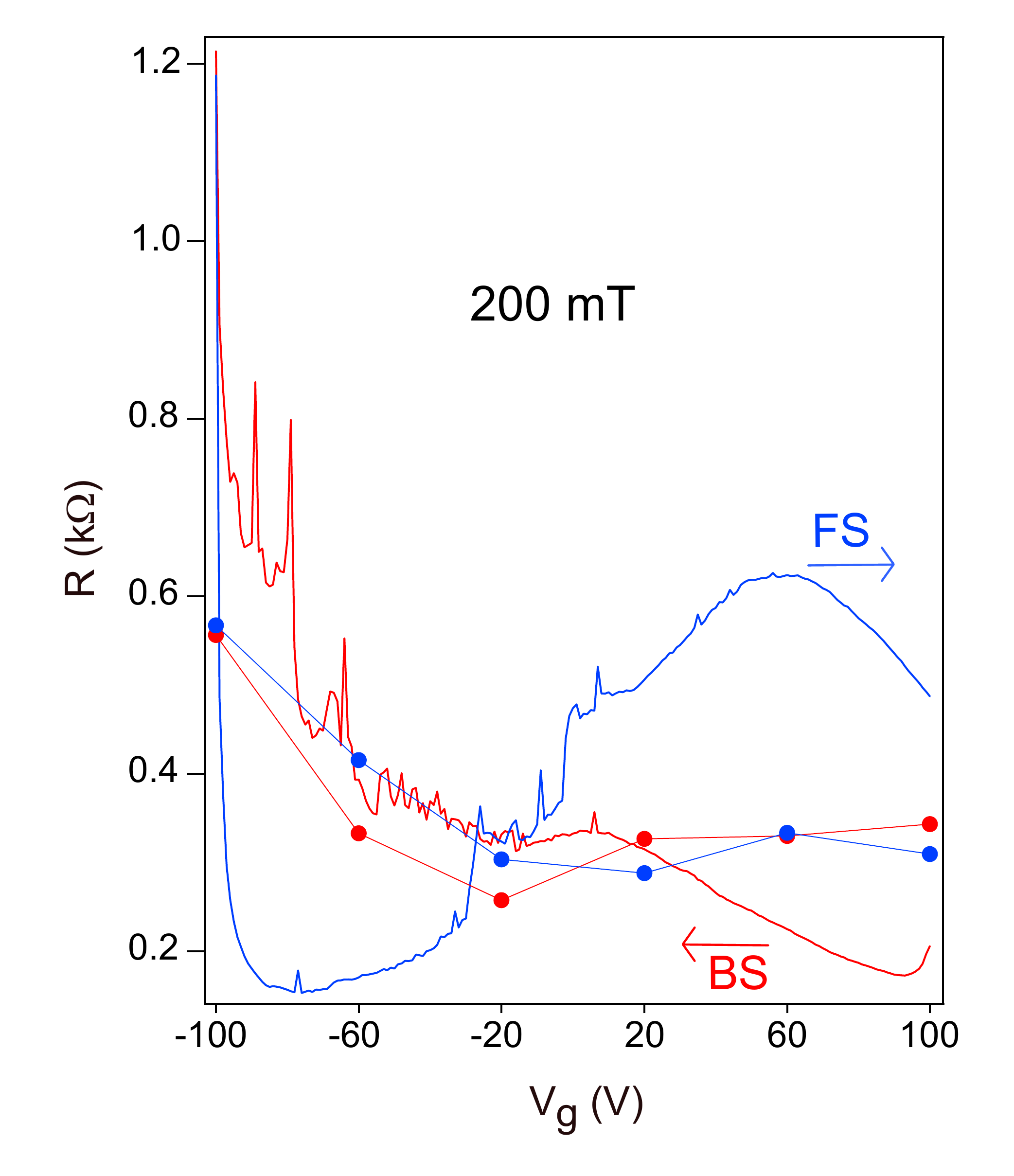}
\caption{From Figs. 4a and 4b of the main text, we obtain the resistance at 200 mT magnetic field at different values of gate voltage $V_g$. These are shown as dots (red and blue correspond to Figs. 4a and 4b of main text respectively). The continuous lines plot the $R$-$V_g$ curve at 200 mT in Fig. 4c of the main text.}
\end{center}
\end{figure}

Magnetoresistance data for fixed values of $V_g$ have been presented in Figs. 4a and 4b of the main text. We now compare (Fig. S3) the variation of resistance $R$ with $V_g$ at a particular value of magnetic field from this dataset, with $R$-$V_g$ plots where $V_g$ is tuned continuously under a constant magnetic field. The $R$-$V_g$ plots at 200 mT (continuous lines) in Fig. S3 are taken from Fig. 4c of the main text. The variation in resistance is more enhanced for the continuous $V_g$ sweep showing that the superconducting system undergoes a more striking transformation in its electronic properties with continuous variation of $V_g$ than could be anticipated from the magnetoresistance data in Figs. 4a and 4b of the manuscript.

\begin{figure}[h]
\begin{center}
\includegraphics[width=80mm]{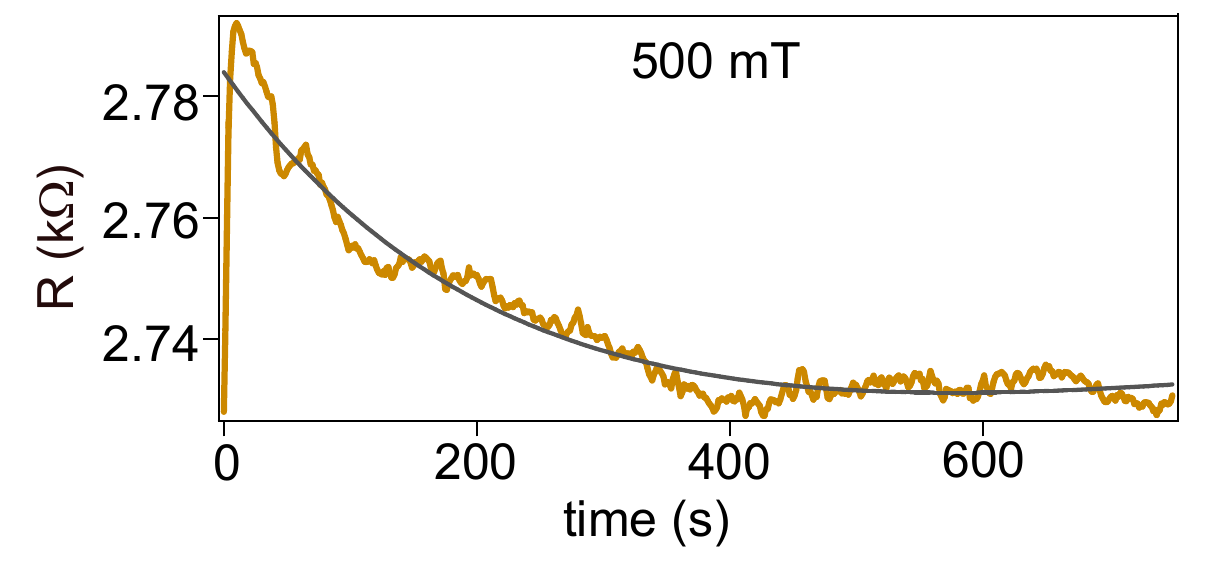}
\caption{This plot is reproduced from Fig. 4d of the main text for clarity. It shows the dynamic change of resistance $R$ with time $t$ at 500 mT after the gate voltage sweep is stopped at -95 V.}
\end{center}
\end{figure}

\begin{figure}[h]
\begin{center}
\includegraphics[width=100mm]{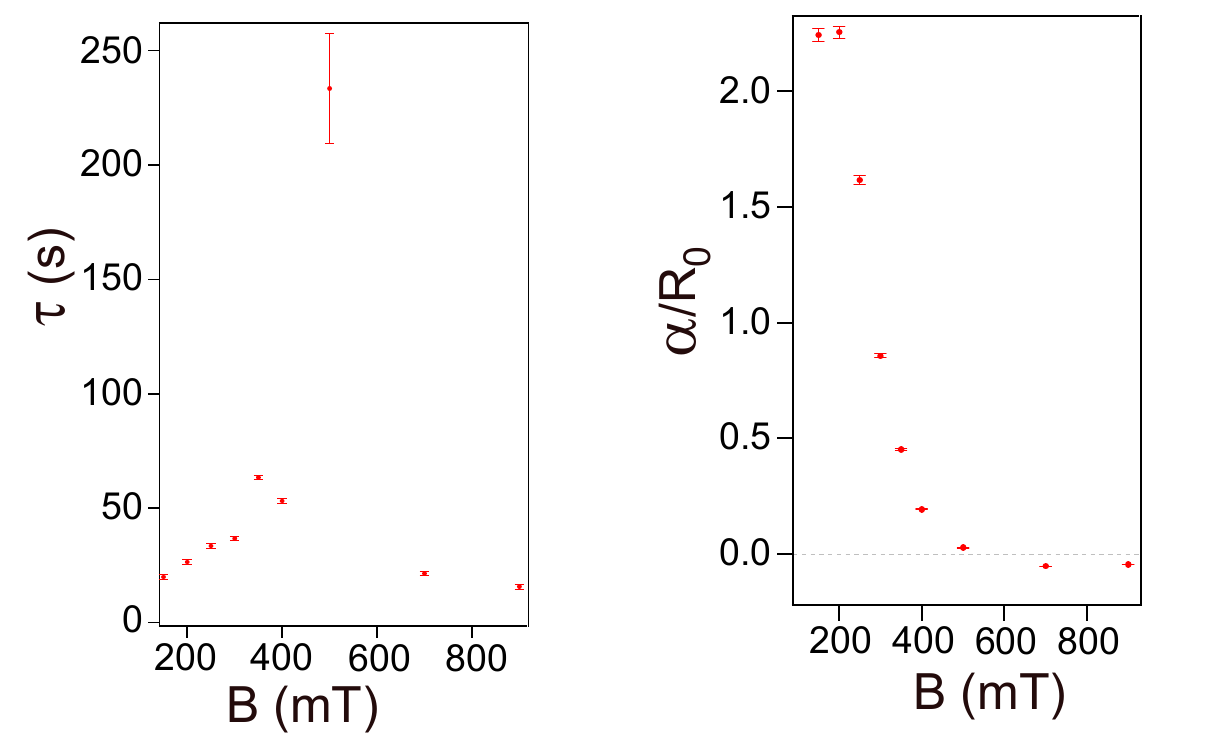}
\caption{The estimated values of $\tau$ and $\alpha/R_0$ from the dataset in Fig. 4d of main text.}
\end{center}
\end{figure}

\section{3. D\lowercase{ynamic change of resistance at different magnetic fields}}

The dynamic change of resistance $R$ of the 2DEG following a sweep of the gate voltage was studied at different magnetic fields and the data was presented in Fig. 4d of the main text. The data was fitted to the function $R(t)=R_0 +$ $\alpha$~$e^{-t/\tau}$ $+$ $\beta$~$t$. $\tau$ is the characteristic time over which the sharp change in $R$ occurs once the gate voltage value is stabilized at -95 V. $\alpha$ is the magnitude of this drop (with the fractional change given by $\alpha/R_0$). As the magnetic field is progressively reduced starting from 900 mT, a huge increase in $\tau$ is observed at 500 mT. The plot showing the dynamic change of $R$ at 500 mT is shown in Fig. S4. This is the same plot as shown in Fig. 4d of main text - it is reproduced here for clarity. $R$ continues to increase for the initial 10 seconds, after which the trend reverses and a gradual reduction in $R$ begins. The estimated values of $\tau$ and $\alpha/R_0$ for the different values of magnetic field (from the dataset in Fig. 4d of main text) are shown in Fig. S5.

\end{document}